\documentclass[aip, amsmath, amssymb, twocolumn, a4paper, 10pt]{revtex4-2} 

\usepackage{graphicx}
\graphicspath{{./}}
\usepackage{dcolumn}
\usepackage{siunitx}                      
\usepackage{bm}

\usepackage[utf8]{inputenc}
\usepackage[T1]{fontenc}
\usepackage{mathptmx}

\usepackage{xcolor}

\newcommand{\Fp}{\ensuremath{F_\text{p}}}
\newcommand{\NAOF}{NA$_\text{S}$}
\newcommand{\NAU}{NA$_\text{C}$}
\newcommand{\TPN}{N$_\text{ph}$}
\newcommand{\NDBR}{N$_\text{DBR}$}

\begin{document}

\title[Hybrid Mie-Tamm photonic structure]{Design of a hybrid Mie-Tamm photonic structure as a highly directional GHz single-photon source}

\author{J. M. Llorens}
 \email{jose.llorens@csic.es.}
\affiliation{Instituto de Micro y Nanotecnología, IMN-CNM, CSIC (CEI UAM+CSIC) Isaac Newton, 8, E-28760, Tres Cantos, Madrid, Spain}
\author{B. Alén}%
\affiliation{Instituto de Micro y Nanotecnología, IMN-CNM, CSIC (CEI UAM+CSIC) Isaac Newton, 8, E-28760, Tres Cantos, Madrid, Spain}

\date{\today}

\begin{abstract}
We present a photonic structure where Mie, Tamm, and surface plasmon optical modes can be tailored to enhance the brightness of an embedded single-photon emitter. Contrary to most proposals, the structure is designed for excitation and collection through the substrate side. The front surface can be used instead to arrange metal contacts which serve both, as electrical gates and optical mirrors. The design is particularized for InGaAs QDs on GaAs resulting in an outcoupled single-photon rate exceeding 3 GHz in a narrow cone of NA = 0.17. The fabrication tolerances are also discussed.
\end{abstract}

\maketitle

\section{\label{sec:Intro} Introduction}

A single-photon source (SPS) with an emission rate beyond 1 GHz is a key element in the field of quantum communications and linear quantum computation~\cite{pirandola_advances_2020, lu_quantum-dot_2021}. 
An ideal SPS requires a near-unity brightness, purity, and indistinguishability.~\cite{senellart_high-performance_2017}. This is particularly relevant for scalable quantum information processing~\cite{wang_boson_2019}.
A major obstacle in obtaining efficient sources is not related to the quantum efficiency of the emission process but to the difficulty in extraction of the emitted photons from a planar surface. The high refractive index of semiconductors defines a narrow solid-angle free of total internal reflection. Taking as an example the GaAs-air interface, roughly 98\% of the emitted photons are trapped within the sample. By placing a planar mirror in close proximity to the emitter, this limit can be raised increasing the extraction from $1/4n^2$ to $1/n^2$, i.e. a factor four.~\cite{benisty1998impact} To further circumvent this limitation, three-dimensional photonic structures are necessary, which either act on the near-field of the emission such as textured surfaces or micropillars, or in its far-field such as solid immersion lenses.~\cite{barnes_solid-state_2002} The extraction problem can be defined as a molding of the wavefront to funnel the light into a solid angle as narrow as possible. If a large extraction is combined with a large Purcell factor (Fp)~\cite{liu_high_2018}, the performance of the source can be increased even further under continuous wave excitation or under pulsed excitation when the excitation repetition rate matches the inverse of the decay rate. This applies particularly in quantum key distribution (QKD), where the secret key rate is proportional to the product of the emission rate and collection efficiency, i.e. total collected photon number {\TPN}~\cite{kupko_tools_2020}.
Micro-pillars and micro-lenses are very successful technologies to increase the brightness of single-photon sources.~\cite{reitzenstein_quantum_2010,ding_-demand_2016} The combination of a high finesses planar distributed-Bragg reflector together with a high aspect ratio of tapered micro-pillars allows for a very smooth extraction of the light from the semiconductor. 
Micro-lenses also offer excellent collection efficiencies~\cite{morozov_metaldielectric_2018, schmidt_deterministically_2020}, but limited {\Fp} mainly due to the large volume compared with the emission wavelength in the semiconductor. 
A better trade-off between collection efficiency and {\Fp} is found in circular Bragg gating (CBG) also known as bull's eye. The best performance reported so far is for upper extraction with a CBG written in a slab surrounded by air resting on SiO$_2$/Ag support.~\cite{wang_-demand_2019,rickert_optimized_2019,kolatschek_bright_2021}

In this work, we will analyze single-photon emitters embedded in a cylinder defined vertically by a bottom distributed Bragg reflector (DBR) and a top metallic mirror, ~\cite{benisty1998impactII} and laterally by dielectric confinement with a low refractive index layer. The aim is to optimize the {\Fp} and collection efficiency of single photons at the substrate-air interface. This way, the top metallic mirror can be used as an electrical contact
to switch the single-photon emitter on and off or tune its properties. Combined with an embedded excitation source underneath, this design could be used to build a monolithic electrically driven and tuneable SPS.~\cite{EP3361516B1} 

\begin{figure*}[ht]
        \centering
        \includegraphics[width=\textwidth]{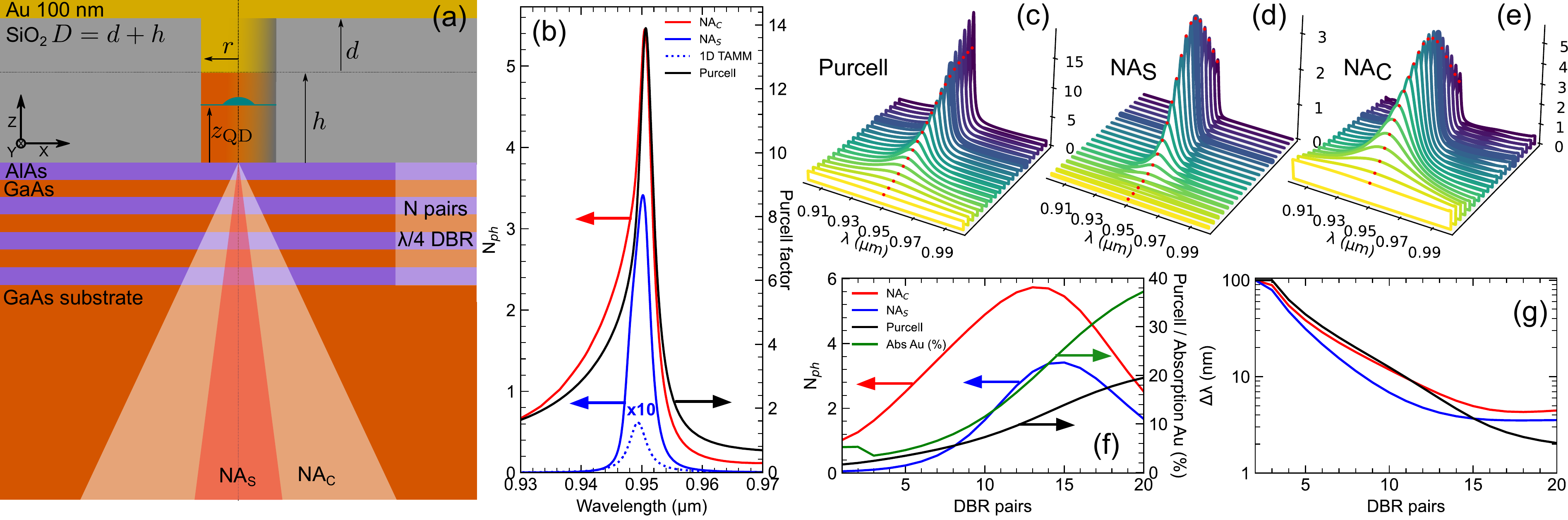}
        \caption{(a) Sketch of the OMTC. (b) Total photon number within GaAs for two different NA, the corresponding {\Fp} at optimal parameters (solid lines). Emission from a planar OTC GaAs microcavity magnified by a factor 10 at {\NAOF} (dashed line) (c-e) Fan plot on the evolution of the emission spectra with {\NDBR} (brightest yellow {\NDBR}=1, darkest purple {\NDBR=20}, bold line {\NDBR=13}). The dot in each line points to the target wavelength (950 nm). Dependence of the peak's maximum (f) and linewidth (g) of the emission for {\NAU}, {\NAOF}, {\Fp}, and absorption in Au.}
        \label{fig:main}
\end{figure*}

Planar asymmetric cavities harness optical Tamm states (OTS) and can exhibit very high Q values~\cite{kaliteevski_tamm_2007,zhou_multiple_2010,sasin_tamm_2008,symonds_high_2017}. However, their planar geometry does not provide lateral confinement, a limitation that has been partially alleviated in the past by reducing the radial extension of the metallic layer.~\cite{gazzano_evidence_2011,braun_enhanced_2015,parker_telecommunication_2019} Increased performance can be obtained by substituting the homogeneous cavity medium with a low refractive index dielectric layer embedding a high refractive index cylinder, as explained below. As a result, Mie-type resonances appear that enhance the electric field~\cite{bidault_dielectric_2019}. Furthermore, the overlap of such Mie resonances in the cylinder with the OTSs of the embedding dielectric layer increases the {\Fp} and steers the emitted light into highly directional cones of light. Surface plasmon resonances near the metallic layer also play a role and shall be taken into account to avoid non-radiative losses. The sketch presented in Figure \ref{fig:main} graphically depicts the four main parts of the optical Mie-Tamm cavity (OMTC) just described. From top to bottom, we find the metallic layer, the dielectric layer, the cylinder that embeds the quantum emitter~\cite{schneider_numerical_2018}, and the $\lambda/4$ DBR. With these elements, the emission is steered towards the substrate where an anti-reflecting coating prevents back reflections. At this point, an optical fiber could be applied directly or using secondary optics. Its design is left out of the current study. 

This design can be adapted to different quantum emitters, like NV centers in diamond~\cite{riedel_deterministic_2017}, molecules~\cite{geng_tamm_2019} or 2D materials~\cite{lundt_room-temperature_2016,zhang_polaritonic_2019} and wavelength combinations. Without loss of generality, we will analyze the optimization of OMTC structures on GaAs embedding single InGaAs self-assembled quantum dots (QDs), SiO$_2$ dielectric layer, and AlAs/GaAs DBR. These QDs show excellent purity~\cite{schweickert_-demand_2018} and indistinguisability~\cite{thoma_exploring_2016}. Moreover, as these quantum emitters are based on III-V semiconductors, they can be integrated into photonic integrated circuits~\cite{hepp_semiconductor_2019,dusanowski_purcell-enhanced_2020,rodt_integrated_2021} and even in silicon substrates~\cite{shang_perspectives_2021}. With these materials choices, we present, in what follows, the optimum design parameters to extract from the sample 3.6 GHz single-photon rates in a narrow cone of NA = 0.17. 

The paper is organized as follows. In section II, we describe the methods used to find the optimum parameters for our design. Section III presents the optimization results, the main properties of the OMTC structure, and some design rules derived from the trade-off between fabrication complexity and device performance. In section IV, we discuss in more detail the nature of the modes harnessed by the structure.

\section{Methods}

The optical cavity is sketched in Fig~\ref{fig:main}(a). The parameters are defined as follows. The top Au mirror thickness is set to 100 nm to reduce losses in the upper space and to reach the saturation region of the phase change~\cite{adams_model_2019}. The dielectric layer thickness is defined as $D=h+d$, where $h$ is the cylinder height and $d$ is an auxiliary parameter. The gap between the GaAs cylinder and the Au mirror is assumed to be occupied by an Au cylinder. Hence, GaAs and Au cylinders take the same radius ($r$) value. The QD is modeled as a dipole emitter perpendicularly oriented to the cylinder axis. It is located in its axis at a distance $z_\text{QD}$. OTSs tend to maximize the field at the mirror interfaces where also maximum enhancement of the emission might take place. However, a QD placed close to these interfaces might be either difficult to grow by self-assembly (DBR side) or suffer from spectral diffusion broadening and/or luminescence bleach (Au-GaAs side)~\cite{wang_optical_2004,liu_solid-state_2019}. Two constraints are imposed, $z_\text{QD}>10$ nm, minimum distance to the top DBR surface, and $z_\text{QD}< h-50$ nm, minimum distance to the metallic cylinder. Both constraints are applied to maximize the fabrication feasibility of the device.

We exploit the axial symmetry of the system to speed up the solution of Maxwell's equations, where the dipole emission asymmetry is handled by an expansion in Fourier modes~\cite{schneider_numerical_2018}. Perfect matching layers are added to the boundary of the simulation region. The numerical solution is performed by the finite-element method with the commercially available software-package JCMsuite (ver 4.5) developed by the company JCMwave GmbH. The refractive indexes of intrinsic GaAs and AlAs at low temperature are computed using the dispersion model in Ref.~\onlinecite{gehrsitz_refractive_2000}, SiO$_2$ from Ref.~\onlinecite{Palik1998}, Si$_3$N$_4$ from Ref.~\onlinecite{luke_broadband_2015} and Au from Ref.~\onlinecite{johnson_optical_1972}. 

The many resonances that occur in this kind of system, naturally call for global optimization methods.~\cite{molesky_inverse_2018, melo_multiobjective_2022, guimbao_numerical_2022}
In our case, the optimization is performed in two stages using the monotonic basin hopping method with a 10\% variation of the free parameters: $r$, $h$, $D$, $z_\text{QD}$. In the first stage, the structure is locally optimized from a starting point and taken as a seed. In the second step, the parameters are varied randomly within a prescribed range. Subsequent iterations finish when no improvement is found after a limited number of trials (five in our case). As a local optimizer, we use the BOBYQA algorithm~\cite{powell2009bobyqa} implemented in the NLopt library~\cite{nlopt}. The whole optimization is performed with the library pygmo~\cite{Biscani2020}. The objective function is defined to maximize the emission within a very narrow cone at the target wavelength of $\lambda_0=950$ nm. Without loss of generality, we set the cone aperture to {\NAOF}=0.17, as a representative value of a highly directional emission, i.e. roughly a sixth of the critical angle ({\NAU}=1). As a figure-of-merit (FOM), we use the total photon number ({\TPN}) defined as the collection efficiency (CE) within a {\NAOF} times {\Fp}.  {\TPN} has a straight-forward interpretation. If one assumes a typical exciton emission radiative lifetime of 1 ns in bulk GaAs ({\Fp}=1), the optimized total photon number can be related to an extracted single-photon rate in the GHz range.
CE is computed from the far-field projection within a homogeneous GaAs medium representing the substrate, ($\mathbf{E}(\theta, \phi)$, being $\theta$ and $\phi$ the polar and azimuth angles in spherical coordinates. The out-coupling to air, i.e. beyond the substrate interface, is also considered by introducing a $\lambda/4$ Si$_3$N$_4$ anti-reflecting coating (ARC)~\cite{gregersen_broadband_2016}. The final extraction efficiency and {\TPN} values on air results from the propagation of $\mathbf{E}(\theta, \phi)$ within homogeneous intrinsic GaAs and its transmission through the GaAs/Si$_3$N$_4$/air interfaces calculated by the transfer matrix method for $s$ and $p$ polarized light. Absorption losses are considered through the tabulated complex refractive index of each material, but at the design wavelength do not lead to additional propagation losses.

\section{Results}

By inspection of Figure~\ref{fig:main}(a), we can envisage the two extreme cases of our design. For zero cylinder diameter, the structure is a SiO$_2$ OTC, and for very large diameters it operates as a GaAs-OTC. For cylinder sizes similar to the wavelength, Mie resonances are expected and we can think of an OMTC cavity. Resonances in an OTC follow the same pattern as in the Fabry-Perot (FP) cavity~\cite{zhou_multiple_2010}, i.e. $d_{N}=d_0 + N \lambda/2n$, where $d_0$ is the minimum resonant thickness, $n$ is the refractive index and $N=0, 1,\ldots$ is the FP order. As a seed value in the optimization, we have considered the thickness of the FP $N=1$ in a SiO$_2$-OTC at 950 nm, i.e. $D=460$ nm. Analogously, we set the initial height of the cylinder to the thickness of the same resonance in a GaAs-OTC, i.e. $h=183$ nm. The other parameters are set to arbitrary initial values, $r=200$ nm and $z_\text{QD}=h/2$ and let the algorithm find the optimal values. The optimal structure is found for $r=223$ nm, $h=361$ nm and $D=459$ nm being the dipole located at 304 nm from the cylinder base. The optimal number of DBR pairs ({\NDBR}) is found to be 15 as explained below. Figure~\ref{fig:main}(b) shows the resonances found around 950 nm for {\Fp} and {\TPN} for light collected within {\NAU} and {\NAOF} within GaAs. At resonance, the maximum {\Fp} is 12.4 and the maximum {\TPN} is 5.3 (3.6) for {\NAU} ({\NAOF}), respectively. The lineshape of the spectrum also depends on the NA. At narrow collection angles, it exhibits a Lorentzian lineshape of 3.7 nm full-width at half maximum (FWHM), while, at the full extraction angle the line shape becomes asymmetric with FWHM=5 nm. 

It is worth comparing the value {\TPN}=3.6 found for {\NAOF} with that of two reference structures. In a substrate with a mirror, the collection efficiency is $\approx1/n^2$.~\cite{benisty1998impact} For $\lambda_0=950$ nm, $n_\text{GaAs}=3.46$, and assuming $\Fp\approx1$, we obtain {\TPN}=0.083 for {\NAU} and {\TPN}=0.0024 for {\NAOF}. The second structure is a GaAs-OTC of $h=179$ nm, i.e. the asymmetric cavity described by Benisty et al. in Ref.~\onlinecite{benisty1998impactII}. By placing the dipole at 10 nm from the DBR (lower bound considered in our calculations), a {\Fp} = 1.4 results in the spectrum shown in Fig.~\ref{fig:main}(b) as a dashed line with a maximum {\TPN}=0.061. Our design, introducing the Mie-Tamm hybrid structure, with the QD located at least 50 nm apart of the nearest surface, improves the bare case by a factor 1500, and the perfected GaAs-OTC case by a factor 60. A clear signature of the advantages brought by the OMTC scheme. 

A key element of the design is the DBR, as it is responsible for the OTS. Figures~\ref{fig:main}(c-e) show the evolution with an increasing number of DBR pairs of the {\Fp} and {\TPN} resonances found around 950 nm. The {\Fp} peak intensity increases monotonically in the studied range, finding a maximum value of 20 for 20 DBR pairs [Figure~\ref{fig:main}(f)]. Meanwhile, {\TPN} resonances have two different regimes where the peak intensity first increases up to 14-15 pairs depending on the NA and then decreases with the number of DBR pairs. This is the expected dependence for collection through the substrate side since the transparency of the DBR mirror decreases with {\NDBR}. In such a case, there is a trade-off between the emission and collection enhancement throughout the substrate and the light storage in the cavity, which is also modulated by absorption at the Au gold mirror and other losses. Indeed, at the {\TPN} maximum, the total scattered power is \SI{75}{\percent} (\SI{72}{\percent} in the downward direction) and the Au absorption is \SI{25}{\percent} (NA={\NAOF}). Beyond that point, the absorption increases up to \SI{37}{\percent}, as shown in Figure~\ref{fig:main}(f).

The resonance linewidth evolution is shown in Figure~\ref{fig:main}(g). Again, the {\Fp} linewidth shows a monotonic dependence, continuously narrowing in the studied range, while the {\TPN} linewidth first narrows downs and then saturates beyond $\approx15$ DBR pairs. This linewidth dependence determines how challenging will be the spectral matching between the QD emission peak and the cavity resonance.~\cite{canet-ferrer_purcell_2012} Depending on the fabrication uncertainty, it could be beneficial to reduce {\NDBR} from 15 to 9 to increase the linewidth from 3.7 nm to 8.7 nm. The penalty is a reduction on {\TPN} from 3.6 to 1.2 GHz for {\NAOF}. This trade-off between performance and fabrication yield can be anticipated from our analysis.

\begin{figure}[htb]
        \centering
        \includegraphics[width=\columnwidth]{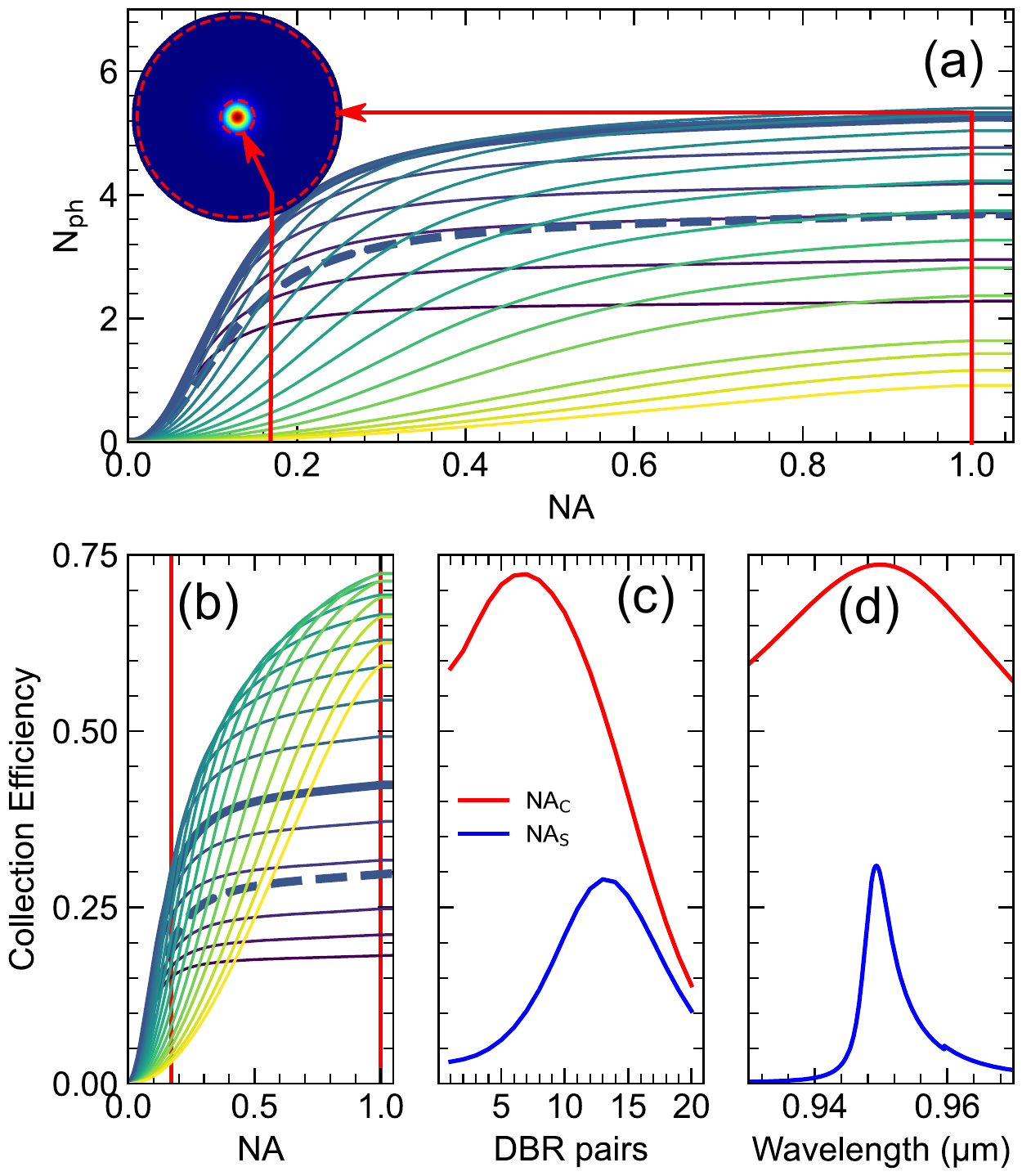}
        \caption{(a) Cumulative {\TPN} in air considering an anti-reflective coating as function of NA for different {\NDBR} (brightest yellow {\NDBR}=1 and darkest purple {\NDBR}=20). Inset, far-field projection up to {\NAU}. Dashed-line circles mark the {\NAOF} and {\NAU} values, also indicated by red arrows. (b) Equivalent plot for CE. (c) Evolution of CE for {\NAOF} and {\NAU} as a function of {\NDBR}. (d) CE spectral dependence at {\NDBR}=7 for {\NAU} and {\NDBR}=13 for {\NAOF}. The solid (dashed) bold line in (a) and (b) corresponds to the optimal {\NDBR}=15 with (without) anti-reflective coating.}
        \label{fig:CE}
\end{figure}

High directionality is demanded to overcome total internal reflection at the substrate interface and bring the emitted light out of the sample.  Figure~\ref{fig:CE} (a) and (b) shows {\TPN} and collection efficiency (CE) on air as a function of NA for varying {\NDBR}. 
The Figure follows the same color scheme as in Figs.~\ref{fig:main}(c-e). A bolder line is used for {\NDBR}=15. The steep rise of {\TPN} and CE for small values of NA is a clear signature of the high directionality of the emission of our design, most of the light has been emitted for $\mathrm{NA}<0.5$. As shown in the inset by a contour polar plot, the far field is concentrated in a very small NA area and lacks azimuth dependence. Peak values obtained in air are only \SI{1}{\percent} smaller than the values reported in Fig.~\ref{fig:main}, because the transmission is close to one for those angles at which the far field is strong. Beyond NA=1, the light can not escape out of the substrate, and {\TPN} remains constant. The dashed line represents the {\TPN} without the Si$_3$N$_4$ ARC for {\NDBR}=15, showing the importance of this additional layer. Without the ARC the reduction is \SI{30}{\percent}.

In Figure~\ref{fig:CE}(c) the CE values at {\NAU} and {\NAOF} as a function of {\NDBR} are presented. For {\NAU} ({\NAOF}) the maximum CE is 0.72 (0.29) obtained at 7 (13) {\NDBR}. A new trade-off is established depending on which magnitude is maximized. Collection at large NA requires a small {\NDBR}, while for narrow NA, a large value of {\NDBR} maximizes both CE and {\TPN}. 
The spectral dependence is shown in Fig.~\ref{fig:CE}(d) for the 7 (13) {\NDBR} at {\NAU} ({\NAOF}), i.e. at the maxima of Fig.~\ref{fig:CE}(c). While CE at {\NAOF} critically depends on the tuning between resonances, CE greater than 0.6 can be obtained at {\NAU} over a much broader band. In summary, the target application and the collection setup will ultimately determine the optimal {\NDBR}.

\begin{figure}[htb]
        \centering
        \includegraphics[width=\columnwidth]{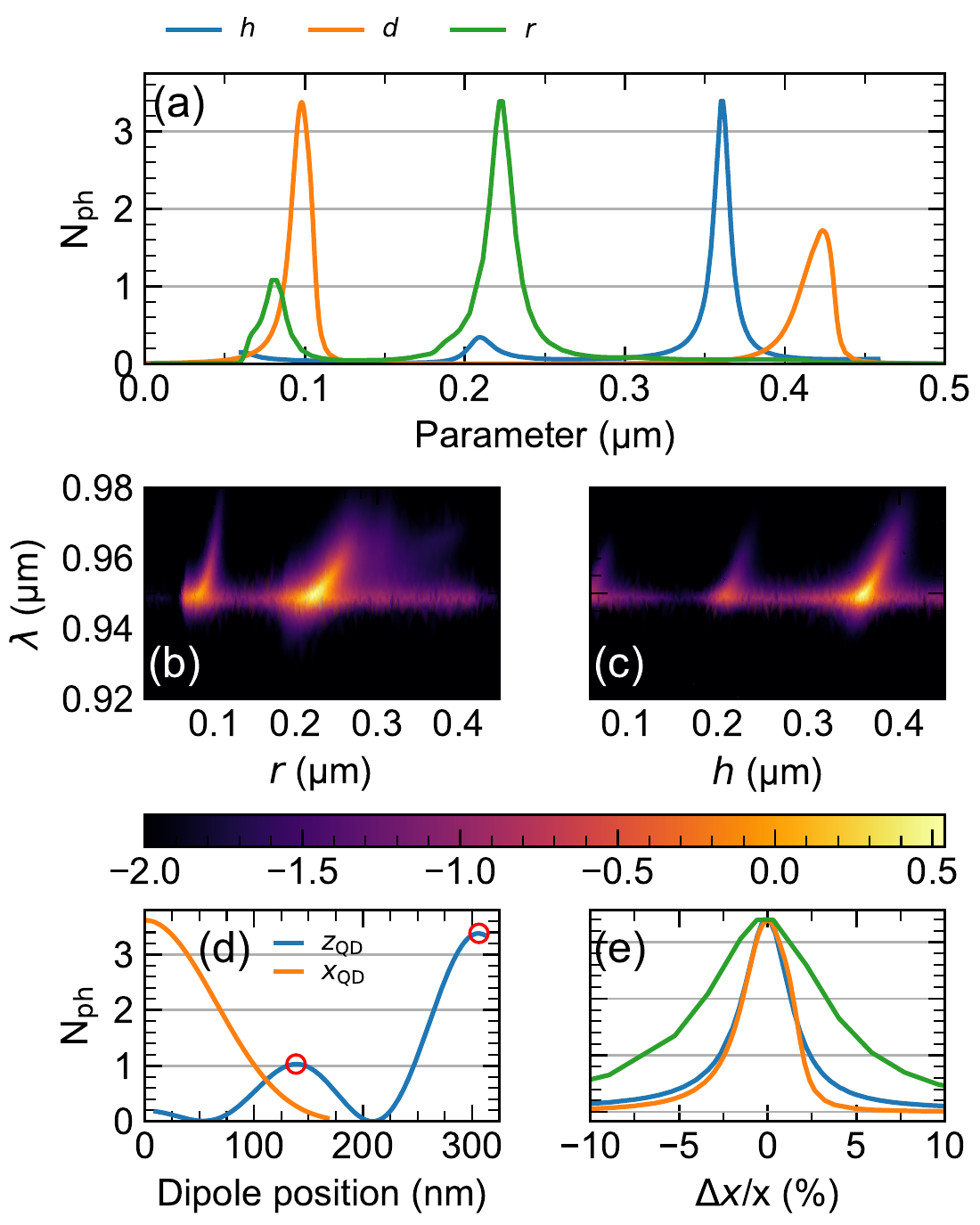}
\caption{Dependence of {\TPN} on the cylinder height $h$, radius $r$, and dielectric thickness relative to height $d$, as calculated for {\NAOF} within GaAs. {\TPN} finds its maximum at $r=223$ nm, $h=361$ nm, and $d=98$ nm. (a). Log plot of {\TPN} at {\NAOF} as function of $\lambda$ and $r$ (b), and $\lambda$ and $h$ (c). (d) Position of the dipole in the cylinder ($x_\text{QD}$ and $z_\text{QD}$) The two maxima in $z_\text{QD}$ discussed in the text are indicated by an open circle. (e) Relative deviation with respect to the optimal value.} \label{fig:params}
\end{figure}

To understand how the optical mode structure varies with the geometrical parameters, we depict in Figure~\ref{fig:params} the dependence of {\TPN} on either $d$, $h$ or $r$, while keeping the other two parameters and $z_\text{QD}$ fixed. The studied range is limited to 500 nm. The SiO$_2$ slab thickness $d$ shows two main resonances at 423 nm and 98 nm [Fig.~\ref{fig:params}(a)]. The free spectral range (FSR) is 325 nm, matching very well $\lambda_0/2n$ in SiO$_2$ ~\cite{llorens_absorption_2014,zhou_multiple_2010}.
The GaAs cylinder height $h$ exhibits also two {\TPN} resonances in Figure~\ref{fig:params}(a), being the FSR=151 nm. This value differs by 14 nm from $\lambda_0/2n$ in GaAs, as expected given the limited in-plane extension of the nanocylinder. Meanwhile, $r$ shows also two resonances, being the FSR=142 nm. The three-dimensional nature of the nanocavity makes their study more involved. Indeed, we study the interplay between $h$ and $r$ in more detail in Section~\ref{sec:Discussion}.

This parameter exploration is done at fixed $\lambda_0$. To get a more comprehensive picture, spectral dependence can be added to the analysis. Figs.~\ref{fig:params}(b,c) show contour plots of {\TPN} in logarithmic scale for different values of $r$ and $h$ and emission wavelengths. The emission is negligible except at the design wavelength, $\lambda_0$, where the enhancement takes place at particular $r$ and $h$ values giving rise to the resonances discussed above. Due to an exponential decay of {\TPN} as soon as the cylinder size and wavelength depart from the optimal values, changing the cylinder radius alone does not provide an effective way to spectrally tune the OMTC and the QD in a broad range. Broad spectral tuning can still be done after growth, within the stop band of the DBR ($\sim$100 nm), reducing $h$ through etching as necessary in order to target smaller $r$ and $\lambda$. Alternatively, {\NDBR} can be reduced as explained above to relax the spectral matching condition.

\begin{figure}[htb]
        \centering
        \includegraphics[width=\columnwidth]{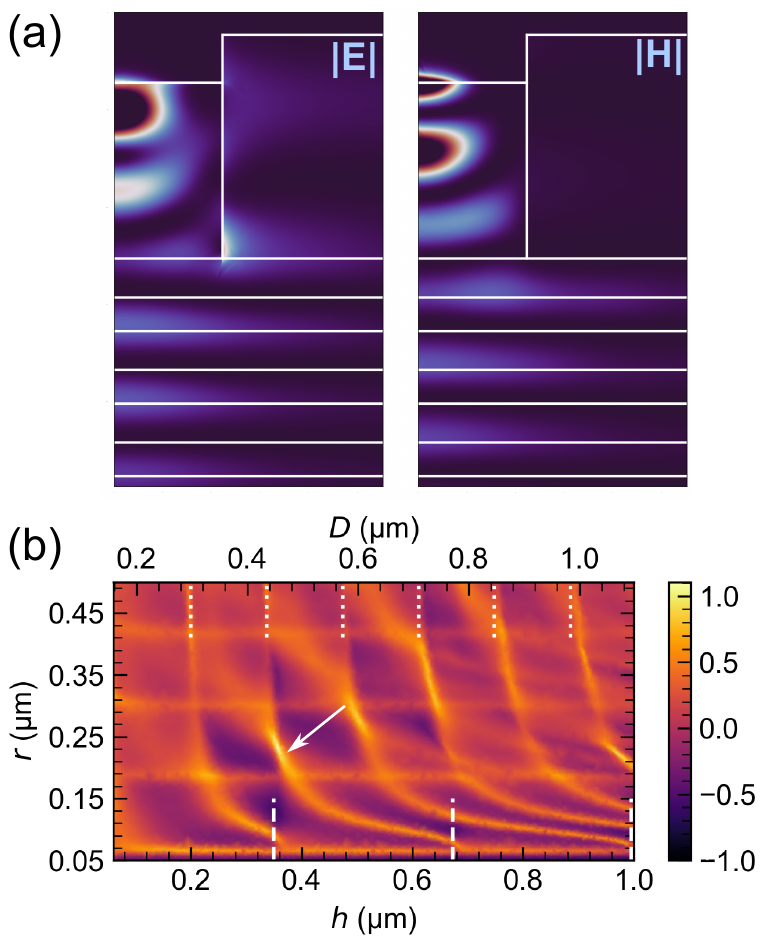}
        \caption{(a) $XZ$ cross-section of $|E|$ and $|H|$ of the optimal structure. The structure layout is defined by the white lines. (b) Dependence of {\Fp} in log scale at 950 nm as a function of $r$ and $h$ for a fixed $d$. The vertical lines denote the operating wavelength in GaAs (dotted lines) and in SiO$_2$ (dashed lines).} 
        \label{fig:RH_fields}
\end{figure}

The last parameters to be discussed are $z_\text{QD}$ and $x_\text{QD}$, the QD off-axis deviation. As shown in Figure~\ref{fig:params}(d) for the optimal geometry, {\TPN} exhibits two maxima at $z_\text{QD}$ 138 and 306 nm. They are rather broad with FWHMs along the cylinder axis of 76 and 47 nm, respectively. They stem from the overlap with the field distributions shown in Fig.~\ref{fig:RH_fields}(a) which, as discussed in the next section, correspond to the cylinder Mie resonance of the OMTC structure. These field distributions tend to occupy a large portion of the nanocavity volume which is highly beneficial for the spatial matching of the single photon emitter and the cavity mode. As it can be seen in Figure~\ref{fig:params}(d),  $x_\text{QD}$ peaks the origin with an FWHM of 149 nm well above the typical precision $\sim$40 nm of deterministic nucleation site-control methods.~\cite{martin-sanchez_single_2009, pregnolato_deterministic_2020, rodt_high-performance_2021} 

The results just presented can be used to discuss the fabrication feasibility of the proposed design. Figure~\ref{fig:params}(e) shows the {\TPN} evolution upon a change of the three most critical parameters around its optimum value. $h$ and $d$ are the most demanding with \SI{3.3}{\percent} and \SI{4.4}{\percent} tolerance each. This stems from resonance FHWMs of 12 nm and 16 nm, respectively. The less demanding parameter is $r$ exhibiting FWHM (tolerance) of 18 nm (\SI{8.2}{\percent}).
Reported experimental values of statistical variations of e-beam lithographic micropillars (30 nm on \SI{2.0}{\micro\meter} diameter) and nanocylinders in metasurfaces (5 nm in 114 nm diameter) can be as low as \SI{1.5}{\percent}~\cite{heuser_fabrication_2018}  and \SI{4}{\percent}~\cite{patoux_challenges_2021}, respectively. These values are below the required tolerance of $r$ in our case. Controlling the SiO$_2$ thickness within tolerance can be done by optically monitored dry etching to reach the target value from an overgrown sample. The results on Figure~\ref{fig:params}(e) indicate that even with deviations twice as large, {\TPN} can reach 0.8 within {\NAOF} (or 800 MHz single-photon count rates extracted from the sample).

Depending on the aspect ratio of the structure, the surface roughness has a great impact on the optical properties. In the case of nanopillars with large aspect ratios, diffraction at the wall roughness produces scattering of the guided mode into other modes resulting in a reduction of the Q-factor~\cite{reitzenstein_quantum_2010}. As the height of the micro-pillar decreases and the diameter increases towards a shallow cylinder, the FSR between adjacent Mie resonances becomes larger and hence a weaker scattering by the surface roughness is expected. Similarly, the exciton linewidth can be affected by surface states and traps at a rough surface. Liu et al.~\cite{liu_single_2018} performed a study of these issues on QD-based photonic nanostructures including surface passivation with Al$_2$O$_3$. They showed that the quantum efficiency is not affected at all in the proximity of dry-etched surfaces 50 to 300 nm away from the QD. At 150 nm the peak broadens by a factor 1.38 and, at 300 nm, the surface does not affect the linewidth anymore. In the whole range, the $g^{(2)}(0)$ value remains constant. In our case, we find the optimal $r$=223 nm, and thus a broadening factor of the order of 1.2 in the exciton emission line could be expected. As a final remark, the 55 nm separation between the QD and the metallic mirror might compromise the indistinguishability of the single-photons, though being larger than typical distances employed in plasmon antennas for quantum applications~\cite{koenderink_single-photon_2017,hughes_theory_2019}. This issue can be solved in the current design by moving the QD to the secondary maximum at 138 nm (223 nm below the interface) [see Fig.~\ref{fig:params}(d)] trading indistinguishability by brightness. The expected change in CE is from \SI{29}{\percent} to \SI{24.2}{\percent} In future designs, physical and geometrical parameters of the structure can be modified to increase that separation. 
From this discussion, we conclude that OMTC designs like the one presented here are feasible from the fabrication point of view.

\section{Discussion}
\label{sec:Discussion}

The interplay between the OTS and the Mie resonances can be inferred from the field distribution shown for the optimized structure in Figure~\ref{fig:RH_fields}(a). The white lines indicate the different regions depicted in Fig.~\ref{fig:main}(a). In the SiO$_2$ slab (top-right) a node line spans parallel to the substrate. This corresponds to the OTS resonance used as seed in the optimization, i.e. a full oscillation of $\lambda_0/2$ in SiO$_2$ fits within the structure. In the DBR region (bottom half region) an oscillation of the electric field in phase with the DBR period is found. Indeed, despite the three-dimensional character of the cylinder, the wavefront is parallel to the DBR's interfaces leading to the narrow directionality of the emission. The field distribution inside of the cylinder clearly shows the excitation of a Mie resonance as anticipated in the discussion of $x_\text{QD}$. In this particular case, $|E|$ shows two ring structures, and $|H|$ shows two anti-nodes. This field distribution is characteristic of a magnetic multipole resonance~\cite{garcia-etxarri_strong_2011,groep_designing_2013,raya_numerical_2019} and has been previously reported for a horizontal dipole located near the top or bottom surfaces in free-standing cylinders~\cite{rocco_controlling_2017}.

The properties of the OMTC cavity are better understood from the evolution of {\Fp} as a function of $r$ and $h$ for fixed $d$, i.e. $D$ changes with $h$.  Figure~\ref{fig:RH_fields}(b) shows the corresponding contour plot with all the resonances excited in the system. The white arrow points to the optimal one. We recall that $r$ determines whether the structure behaves as a SiO$_2$-OTC ($r\ll$) or as a GaAs-OTC ($r\gg$). For $r=50$ nm discrete resonances appear with an FSR of $\lambda_0/2$ corresponding to the SiO$_2$-OTC and indicated by vertical white dashed lines. 
At $r=450$ nm, another family of resonances appears, with an FSR of $\lambda_0/2$, which is associated with the GaAs-OTC by a series of vertical white dotted lines. The modal structure of the cylinder offers enough flexibility to smoothly connect both limits. The optimal structure appears when $h$ gets close to the GaAs-OTC FP $N=2$. Around this $h$, the SiO$_2$-OTC FP $N=1$ mode is also excited. This explains why $|E|$ exhibits two nodes in the cylinder while only one in the SiO$_2$ slab in Fig.~\ref{fig:RH_fields}(a). The Purcell factor contour plot also reveals a series of horizontal resonances crossing the Mie resonances at regular $r$ intervals. They are related to the excitation of surface plasmon polaritons at the GaAs/Au interface as explained below.

\begin{figure}[htb]
        \centering
        \includegraphics[width=\columnwidth]{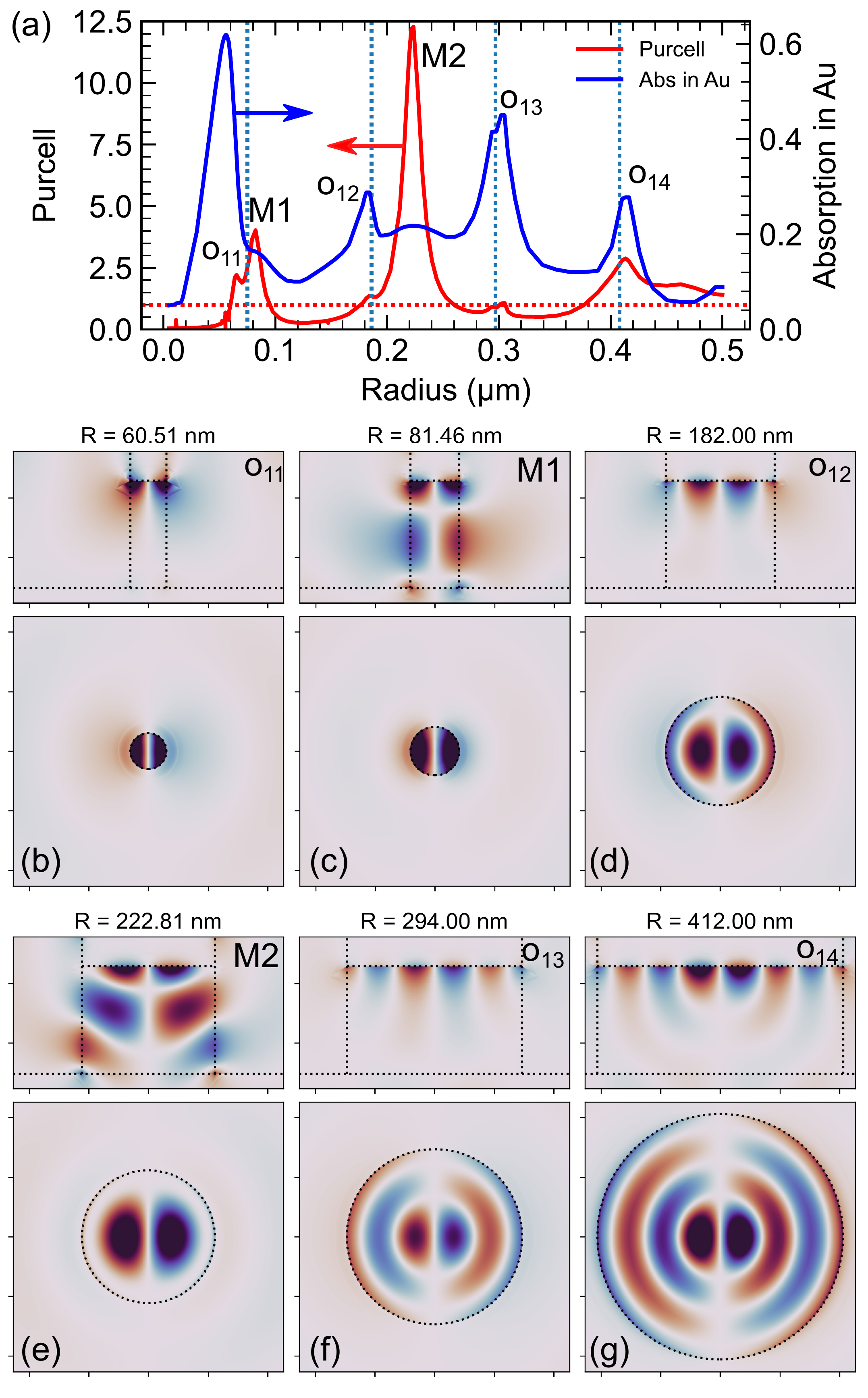}
        \caption{(a) Evolution of {\Fp} and absorption in Au as function of $r$ for the optimal $h$. The horizontal red dotted line indicates unity {\Fp}. The vertical blue dotted lines are spaced by $\lambda_\text{SPP}/2$ (see main text). (b-g) Top: contour plot of $\Re{(E_z)}$ in an XZ cross-section centered at the origin (dotted lines are the profile of the structure) Bottom: Analogous XY cross-section close to the GaAs/Au interface. All plots use the same color scale between negative (red) and positive (blue) field values.
        }
        \label{fig:Plasmon}
\end{figure}

 Figure~\ref{fig:Plasmon}(a) shows a line plot of {\Fp} as a function of $r$ fixing $h$ at the optimal value. The blue line stems from the absorption in the Au mirror calculated as the fraction of the total emitted power as explained above. We distinguish two series of mode peaks labeled O$_{1n}$ and M$_n$, respectively. Their different nature is revealed by plotting in Figs.~\ref{fig:Plasmon}(b-g) the cross-sections of $\Re{(E_y)}$ in the XY plane (top) and XZ plane (bottom) for each mode. The field distributions in (b), (d), (f), and (g) are clearly confined in the GaAs/Au interface, while those in (c) and (e) occupy the volume of the GaAs cylinder. In view of these results, the latter are attributed to Mie resonances of different magnetic multipole order (M$_1$ and M$_2$) while the former can be related to surface plasmon polaritons (SPPs).

Our analysis reveals that, in the region of the GaAs/Au interface, the metallic contact acts as a circular plasmonic patch nanoantenna.~\cite{filter_circular_2012}. In our configuration, the dipole is oriented parallel to the interface and excites the family of odd modes labeled as O$_{mn}$~\cite{minkowski_resonant_2014}. In particular, the modes with $m=1$, the first azimuth mode number, and different $n$ radial ones. The FSR between the O$_{1n}$ peaks [dotted vertical lines in Fig.~\ref{fig:Plasmon}(a)] compares very well to half-wavelength of the SPP $\lambda_\text{SPP} = \lambda_0 \sqrt{(\Re(\varepsilon_\text{Au})+\varepsilon_\text{GaAs})/\Re(\varepsilon_\text{Au}) \varepsilon_\text{GaAs}}$~\cite{novotny2006principles} being equal to 109 nm. It also agrees very well with the resonance condition of circular patch nanoantennas~\cite{filter_circular_2012}:
\begin{equation}
        2\Re(k_\text{SPP})r_{n} + \phi^r = 2x_n,
\end{equation}
where $r_{n}$ is the $n$-th radius at resonance, $\phi^r$ is the phase origin, $k_\text{SPP}$ is the SPP wavevector, and $x_n$ is the $n$-th zero of the $J_1$ Bessel function. The phase $\phi^r$ depends on $r$, but, as a first approximation, we can assume that the dependence is very weak. Hence, $r_{n+1}-r_n=(x_{n+1}-x_{n})/\pi \lambda_\text{SPP}/2$. In addition, verifies that $(x_{n+1} -x_n)\approx\pi$ for small $n$, and therefore the FSR is very close to $\lambda_\text{SPP}/2$, as expected. This is the final confirmation of the nature and position of the different SPP resonances found in our system.

One needs to be aware of the presence of these SPP resonances when designing an OMTC structure. They can introduce large parasitic losses jeopardizing the enhancement of {\Fp} and {\TPN} produced by the combination of Mie and OTS resonances. So, even if our design is fully scalable to other wavelengths, e.g. \SI{1.3}{\um} or \SI{1.5}{\um} just by rescaling the DBR, $r$, $h$ and $D$ accordingly, some care needs to be taken in case of the material dispersion introduces the unwanted overlap between SPPs and Mie-resonances just described. Otherwise, a penalty in performance would be paid.

We finish this Section by reviewing the main assets of our approach. The backwards collection through the substrate side allows to use a metallic optical mirror and ohmic contact to drive and tune electrically the emission. The cylinder height of 350 nm allows to fit a \emph{p}-\emph{i}-\emph{n} diode without problems. If driving is not necessary and fine control of the charge and wavelength is desired instead, an ultra-thin \emph{p}-\emph{i}-\emph{n}-\emph{i}-\emph{n} diode of 162.5 nm~\cite{lobl_narrow_2017} or a simpler Schottky diode ~\cite{alen_continuum_2005} could be used. In all cases,  QDs to surface distances can be adapted and values of 100-250 nm are typical. The doping layers will introduce some additional losses, without a significant impact on the expected device performance. Representatives values of the absorption in GaAs at $\lambda\approx$\SI{1}{\micro\meter} are $\alpha=$11 (5) cm$^{-1}$ per 10$^{18}$ cm$^{-3}$ \emph{p}-type (\emph{n}-type) doping~\cite{hegblom1999engineering}, and hence the refractive index imaginary part of the doped regions would result in 8$\times$10$^{-5}$ (4$\times$10$^{-5}$). The design relies on a semitransparent DBR located underneath the QD, which enables optical excitation through a built-in~\cite{lee_electrically_2017, munnelly_electrically_2017} or external laser diode. Secondary micro-optical elements can be located at the substrate to excite and/or collect the photoluminescence, which would not interfere spatially with the primary photonic nanocavity embedding the QD. Compared with photonic crystal microcavities, the removal of material surrounding the cylinder reduces the chances of unwanted luminescence from nearby QDs contributing to the optical mode. Also, the Mie resonance in-plane extension fills a great ratio of the cavity cross-section increasing the chances of good spatial overlap in low-density QD samples without requiring deterministic methods. From the performance point of view, the directionality of the emission is superb and allows large extraction single-photon rates within low NA even with moderate Purcell factor values. In addition, the design is fully scalable and manufacturable with current microplanar fabrication techniques.

\section{Conclusions}

We introduce and analyze a photonic structure to enhance the brightness of single-photon sources with very high efficiency. Via a global optimization method, we find the optimal structure parameters for the case of InGaAs QDs embedded in GaAs. From the analysis of the resulting structure, we show that the enhancement is based on the cooperation of Tamm and Mie resonances intertwined in our design. We predict a {\TPN} of 3.6 for a narrow extraction cone of NA$=0.17$ and 5.3 for the critical angle extraction cone (NA$=1$) which boosts the extracted single-photon rate to the GHz range. These values are significantly larger than those attainable in a bare optical Tamm cavity or a bare Mie structure of comparable size. Therefore we term our proposal as an Optical Mie-Tamm Cavity (OMTC). 

\begin{acknowledgments}

This work is supported by European Union’s Horizon 2020 research and innovation programme under the Marie Skłodowska-Curie grant agreement No 956548, project Quantimony,
this project (20FUN05 SEQUME) have received funding from the EMPIR programme co-financed by the Participating States and from the European Union’s Horizon 2020 research and innovation programme, the Agencia Estatal de Investigación (AEI) grant PID2019-106088RB-C31, and CSIC Interdisciplinary Thematic Platform (PTI+) on Quantum Technologies (PTI-QTEP+).

\end{acknowledgments}

\bibliography{main}

\end{document}